\documentclass[preprint]{aastex63}

\submitjournal{ApJ}
\shorttitle{Non-linear mechanisms}
\shortauthors{Jie Jiang}

\begin{document}

\title{Non-linear mechanisms that regulate the solar cycle amplitude}
\correspondingauthor{Jie Jiang}
\email{jiejiang@buaa.edu.cn}

\author[0000-0001-5002-0577]{Jie Jiang}
\affiliation{School of Space and Environment, Beihang University, Beijing, China}
\affiliation{Key Laboratory of Space Environment Monitoring and Information Processing of MIIT, Beijing, China}

\begin{abstract}
The solar magnetic activity cycle has an amplitude that varies within a wide but limited range of values.  This implies that there are non-linear mechanisms that prevent runaway solutions. The purpose of this paper is to propose observable non-linear mechanisms in the framework of the Babcock-Leighton-type dynamo. Sunspot emergences show systematic properties that strong cycles tend to have higher mean latitudes and lower tilt angle coefficients. We use the surface flux transport model to investigate the effect of these systematic properties on the expected final total dipolar moment, i.e. cancellation plus generation of dipole moment by a whole solar cycle. We demonstrate that the systematic change in latitude has similar nonlinear feedback on the solar cycle (latitudinal quenching) as tilt does (tilt quenching). Both forms of quenching lead to the expected final total dipolar moment being enhanced for weak cycles and being saturated to a nearly constant value for normal and strong cycles. This explains observed long-term solar cycle variability, e.g., the Gnevyshev-Ohl rule, which, in turn, justifies the non-linear mechanisms inherent in the Babcock-Leighton-type dynamo.
\end{abstract}

\section{Introduction} \label{sec:intro}
Solar activity shows an 11 yr (quasi)periodicity with a marked, but limited variability of the cycle amplitudes.  One typical example is the Gnevyshev-Ohl rule or Even-Odd effect \citep{Gnevyshev1948}, which is a pattern of alternating higher and lower than average solar cycle amplitudes observed in the sunspot number record \citep{Charbonneau2007}. Furthermore, there are extended intervals of very low or particularly high activity, which are referred to as grand minima and maxima \citep{Usoskin2017}.

It is generally agreed that a dynamo mechanism is responsible for producing the solar magnetic cycle. The basic concept for the large-scale dynamo involves a cycle during which the poloidal and toroidal fields are mutually generated by one another \citep{Charbonneau2020}. The winding of the poloidal field by differential rotation creates a toroidal field, i.e., $\Omega$-effect. A reversed poloidal field results from the twist in the toroidal field by the Coriolis force owing to the solar rotation. Classical mean-field dynamos \citep{Parker1955,Steenbeck1969} and Babcock-Leighton (BL) dynamos \citep{Babcock1961,Leighton1969} are two popular types of dynamos. Major differences between them concentrate on the poloidal field regeneration mechanism and roles of the sunspot emergence. The former relies on the mean electromotive force associated with small-scale convective turbulence in the bulk of the convection zone. The sunspot emergence is the byproduct of the dynamo. The latter captures the buildup of the surface dipole resulting from the emergence of toroidal fields in the form of tilted sunspot groups (Process I) and their subsequent decay and transport by supergranular diffusion and large-scale surface flows (Process II). The so-called BL mechanism consists of the two processes. The sunspot emergence plays an important role in the BL dynamo. It is an essential link in the dynamo loop and serves as a seed for the toroidal field in the next cycle.

There is strong evidence that the dynamo is of BL type \citep{Wang2009, Kitchatinov2011,Munoz-Jaramillo2013, Jiang2013,Cameron2015}. A critical view on classical mean-field dynamos is given by \cite{Spruit2011}. Here we limit our study in the context of the BL-type dynamo.

The production of the toroidal magnetic field is regarded to be linearly proportional to the poloidal field strength at the previous cycle minimum. This is supported by observations \citep{Hathaway2015, Jiang2018}. The regeneration of the toroidal field from the poloidal field by the BL mechanism is expected to be a non-linear process, which is responsible for the bounding growth of the amplitude and cycle variability. But people still do not have a proper understanding of the subsurface processes although MHD simulations of flux tube emergence have provided much insight into its physics \citep{Fan2009}. The thin flux tube simulation with a given toroidal field strength from the bottom of the convection zone to near the solar surface is a long-held paradigm to understand the flux emergence. A typical result from the simulations is a gradual decrease in sunspot group tilts with increasing field strength in the flux \citep{DSilva1993, Caligari1995}. The kinematic BL-type dynamo simply incorporating the characters of flux emergence from thin flux tube simulations is the major method to understand the solar cycle so far.

The tilt angle of the sunspot group is a crucial component of the poloidal field generated from the toroidal one due to the BL mechanism. A decrease in tilts with increasing field strength implies a non-linear feedback mechanism to modulate the efficiency of the BL mechanism. To mimic the nonlinearity, a widely adopted method is a simple form of algebraic quenching in BL-type dynamos. A more physically realistic treatment of the BL mechanism is the `double ring' \citep{Durney1995, Munoz-Jaramillo2010} and `SpotMaker' algorithms \citep{Miesch2014,Karak2017} in 2D and 3D BL dynamo models, respectively. In these models, simple algebraic quenching nonlinearities tend to lock the system to a stable periodic mode with a fixed amplitude, rather than showing irregular behavior.

With the simple algebraic quenching nonlinearity, people have suggested three kinds of mechanisms to understand solar cycle variability. \cite{Charbonneau2005} showed that the combination of the time-delay effect in BL dynamos and the simple tilt-quenching nonlinearity with a lower threshold on poloidal field production can introduce a well-defined transition to chaos through a sequence of period-doubling bifurcations as the dynamo number is increased. The general idea of this mechanism goes back to \cite{Durney2000}. The rising of the flux tube through the convection zone is buffeted by vigorous turbulence, which works as a stochastic forcing generating a large scatter in the tilt angles around the average \citep{Weber2013}. Hence, the BL mechanism has inherent randomness \citep{Olemskoy2013,Jiang2014b}. The stochastic forcing combined with the $\alpha$-quenching is the major mechanism to understand solar cycle variability during the recent several years \citep{Cameron2017,Karak2017, Lemerle2017,Nagy2017}. The variation of the meridional flow is another mechanism to modulate the solar cycle, e.g., \cite{Rempel2006b,Rempel2007,Nandy2011, Choudhuri2012}.

The identification of the realistic non-linear mechanism that operates on the Sun is a necessity to understand the solar cycle, including cycle variability. The simulated property of the tilt, that is, a decrease in tilts with increasing field strength, as a nonlinearity of the BL mechanism has received observational support by analyzing the historical datasets. \cite{Dasi-Espuig2010} showed that the cycle-averaged tilt angles are anti-correlated with the cycle strength. \cite{Karak2017} suggested that a suppression of the tilt by only 1$^\circ$-2$^\circ$ is sufficient to limit the dynamo growth. This raises the following questions. Is there any other nonlinearity relevant to the BL mechanism? Which one is dominant? What kind of roles do they play in regulating the solar cycle? These questions motivated us to set up this study.

Process II of the BL mechanism, i.e., the evolution of emerged sunspot groups over the surface, can be reproduced by the Surface Flux Transport (SFT) model \citep{Wang1989, Mackay2012, Jiang2014a}. Emerging active regions act as a source of surface magnetic flux. Although the processes of magnetic flux rope formation, buoyant rise, and photospheric emergence (Process I) are not yet fully understood, their results, bipolar magnetic regions appearing in the solar photosphere, are at least directly observable. So we use sunspot datasets combined with the SFT model to pin down the nonlinear mechanisms that modulate the dipole moment generation. The nonlinear mechanisms correspond to the ones of the solar cycle in the framework of the BL-type dynamo.

The paper is organized as follows. In Section 2, we give a brief overview of the systematic properties of sunspot group emergence, based on which we suggest two forms of quenching: latitudinal and tilt quenching. In Section 3, we have a short description of the SFT model. We evaluate the effect of latitudinal and tilt quenching, respectively and together, on the solar cycle in Section 4. Our conclusions are given in Section 5.

\section{Systematic properties of sunspot group emergence} \label{sec:ARproperties}
\citet{Jiang2011} gave a detailed study of the dependence of the statistical properties of sunspot emergence on the cycle strength based on the longest available homogenous datasets. They are Royal Greenwich Observatory (RGO) dataset covering cycles 12-20 for latitudes and areas, Mount Wilson and Kodaikanal datasets covering cycles 15-21 for tilt angles. Here we revisit the systematic properties of sunspot group latitude and tilt angle. There are two minor differences from \citet{Jiang2011}. One is that we here define cycle amplitude $S_n$ as the maximum value of the 13-month smoothed monthly sunspot number in version 2 \citep{Clette2014} during cycle $n$. The other is that we investigate the tilt angle properties using all of the tilt angle data, i.e., cycles 15-21, rather than cycles 15-20 used by \citet{Jiang2011}.

Figure 1a shows that stronger cycles have higher mean latitudes with a high statistical confidence level ($p=0.0054$). This was also demonstrated by \cite{Li2003, Solanki2008, Mandal2017}. The linear fit between $S_n$ and $\lambda_n$ is
\begin{equation}
\lambda_n=12.03+0.015S_n,
\label{eqn:latSn}
\end{equation}
where $\lambda_n$ is the mean emergence latitude of cycle $n$. The error bars correspond to standard deviations of the averaged values. The largest difference between different cycles is 3$^\circ$ (16.25$^\circ$ vs 13.25$^\circ$). The relative differences are within 20\% of each other. We will demonstrate in Section \ref{sec:results} that the difference in latitudes has a large modulation of the solar cycle.

Since the tilt angle has a dependence on the latitude, we remove the effect of the latitude by using a parameter $T_n$. A square root function is used to connect the relation between tilt angle $\alpha$ and latitude $\lambda$ \citep{Cameron2010}, i.e.,
\begin{equation}
\alpha=T_n\sqrt{|\lambda|}.
\label{eq:alpha}
\end{equation}
The cycle-averaged tilt coefficient $T_n$ is calculated by
\begin{equation}
T_n=\frac{\sum_i\sqrt{A_i}\alpha_i}{\sum_i\sqrt{A_i}\sqrt{|\lambda_i|}}.
\label{eq:alpha2}
\end{equation}
Here we weight the tilt angle and latitude using the square root of each sunspot group area, which is equivalent to the angular separation of two polarities \citep{Cameron2010}. Other details can be found in \cite{Cameron2010} and \cite{Jiang2011}.

Figure 1b shows the relationship between the cycle strength $S_n$ and tilt angle coefficient $T_n$. $T_n$ is the arithmetic average of its values from Mount Wilson and Kodaikanal. The error bars correspond to standard deviations of the averages calculated by means of error propagation, where the errors for the mean tilt angle, area and latitude correspond to their standard errors measured by Mount Wilson and Kodaikanal. The correlation between them is -0.7 with a confidence level of $p=0.063$. The linear fit between $T_n$ and $S_n$ is
\begin{equation}
T_n=1.72-0.0021S_n.
\label{eq:TnSn}
\end{equation}

We see from Figure 1b that there is a large uncertainty of each $T_n$ and a weak anti-correlation between $T_n$ and $S_n$. See also \citet{Dasi-Espuig2013,Ivanov2012,McClintock2013}. We argue that the weak correlation between $T_n$ and $S_n$ is due to the large scatter and continuous evolution of tilt during a sunspot group lifetime, introducing the uncertainty in tilt measurements. Our recent analysis of tilt angle data set from Debrecen Photoheliographic Data (DPD) during cycles 21-24 shows the anti-correlation with a reasonable confidence level, especially at the maximum area development of each sunspot group corresponding to a smaller tilt uncertainty (Jiao et al. in prep.). But the DPD tilt angle data set is shorter than the data from Mount Wilson and Kodaikanal. The combination of the different data sets is unwarranted since they have large differences in completeness, continuity, method grouping sunspot groups and so on. So here we just use the result for cycles 15-21 based on the data from Mount Wilson and Kodaikanal.

The differences between $T_n$ values are within 40\% of each other. But the largest difference between cycle-averaged tilt $\alpha_n$ is just 1$^\circ$ (5.33 vs 4.30). Here $T_n$ and $\lambda$ vary with the solar cycle in opposite directions, which causes a minor difference in the mean tilt angle.

\begin{figure}[htbp]
\centering
\includegraphics[scale=0.92]{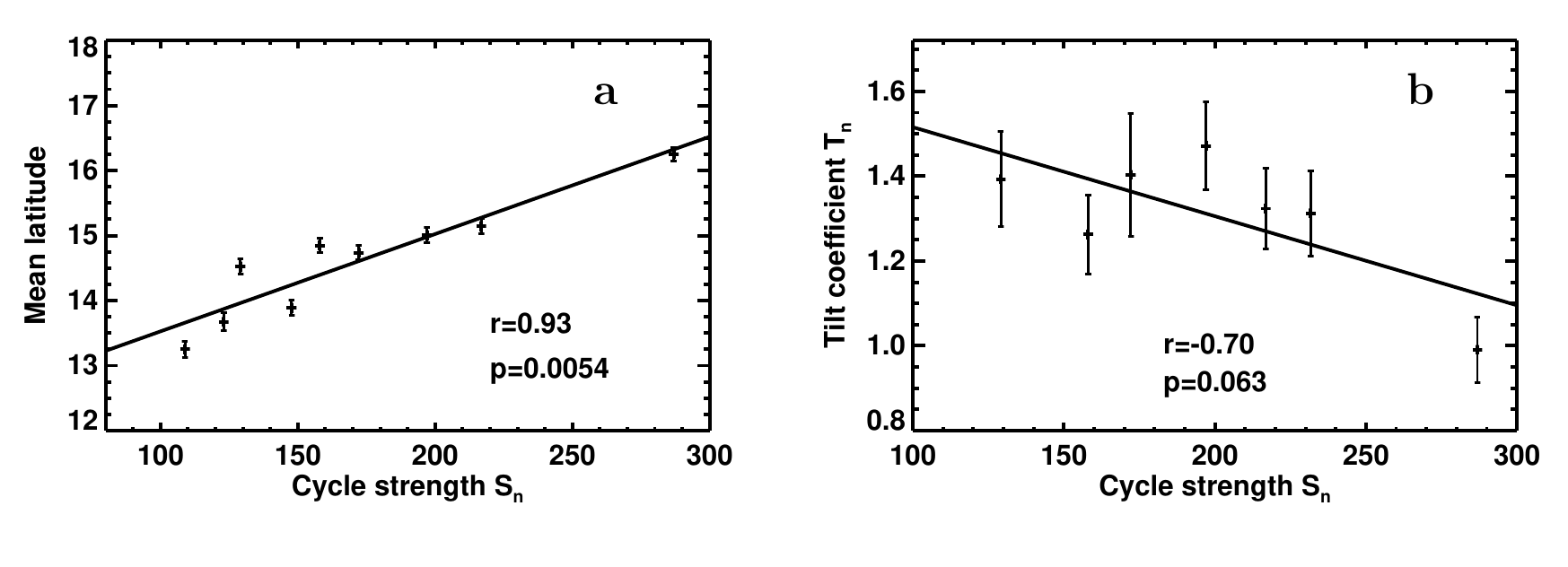}
\caption{Cycle dependence of average sunspot properties. \textbf{a,} Correlation between cycle-averaged latitude of sunspot emergence and cycle strength during cycles 12-20. \textbf{b,} Correlation between cycle-averaged tilt coefficient and cycle strength during cycles 15-21. The error bars represent standard deviations of the averaged values. The lines are linear regressions to the points.}
\label{fig:ARproperties}
\end{figure}

For a strong cycle, a smaller tilt coefficient leads to a weaker contribution to the axial dipole moment and corresponds to a weaker efficiency of poloidal magnetic field production via the BL mechanism. We refer to the effect of the tilt's property as tilt quenching. Higher latitude emergence in strong cycles leads to less flux transport across the equator, which also causes a weaker contribution to the axial dipole moment. The contribution of bipolar sunspot groups to the final surface dipole had recently been shown to be well described by a Gaussian in latitude centered on the equator \citep{Jiang2014b, Nagy2017, Whitbread2018, Petrovay2020}. Lower-latitude emergence contributes much stronger final dipole moment around cycle minimum due to the transport process over the surface if sunspot groups at different latitudes have the same initial axial dipole moments. We refer to the effect of the latitude's property as latitudinal quenching. The effect of the tilt's property on the solar cycle, i.e., tilt quenching, is straightforward. But there are no available publications reporting on the latitudinal quenching. We compare the effects of the two forms of quenching and demonstrate how effective they are in regulating cycle dipole moment generation by performing the SFT simulations in Section \ref{sec:results}, before which we first present the SFT model we use.

\section{Surface Flux Transport model}

In the SFT simulations, the evolution of the radial magnetic field over the surface, $B(\theta, \phi, t)$, is a result of the emergence of sunspot groups, which provides the source of magnetic flux $S(\theta, \phi, t)$, and the subsequent transport of magnetic flux by near-surface plasma flows, including convection, which is usually treated as the turbulent diffusivity $\eta$, differential rotation $\Omega(\theta)$, and poleward meridional flow $\upsilon(\theta)$, where $\theta$, $\phi$ are heliographic colatitude and longitude, respectively. We use the code described in \cite{Baumann2004}. The model solves the radial component of the magnetic induction equation as follows:
\begin{eqnarray}
\label{eqn:SFT} \nonumber
\frac{\partial B}{\partial t}=&-&\Omega(\theta)\frac{\partial
B}{\partial \phi}-\frac{1}{R_\odot\sin\theta}
\frac{\partial}{\partial\theta}\left[\upsilon(\theta)B\sin\theta\right]\nonumber \\
&+&\frac{\eta}{R_\odot^{2}}\left[\frac{1}{\sin\theta}\frac{\partial}{\partial\theta}
\left(\sin\theta\frac{\partial B}{\partial\theta}\right)+
\frac{1}{\sin^{2}\theta}\frac{\partial^{2}B}{\partial\phi^2}\right]\nonumber\\
&+&S(\theta,\phi,t).
\end{eqnarray}

For the surface differential rotation, $\Omega(\theta)$, we use the profile given by \cite{Snodgrass1983}
\begin{equation}
\Omega(\theta)=13.38-2.30\cos^2\theta-1.62\cos^4\theta
\end{equation}
in deg day $^{-1}$. The surface meridional flow, $\upsilon(\theta)$, is described by the profile \citep{vanBallegooijen1998}
\begin{equation}
\upsilon(\theta)=\left\{
  \begin{array}{l l}
     \upsilon_0\sin\left[2.4\ast(90^{\circ}-\theta)\right] & 15^{\circ}< \theta < 165^{\circ} \\
     0 & \textrm{otherwise},
  \end{array}
  \right.
\end{equation}
with $\upsilon_0=11$~ms$^{-1}$. The turbulent diffusivity is taken as $\eta=250\,$km$^2$s$^{-1}$, which is within the range summarized in Table 6.2 of \cite{Schrijver2000}.

For the flux source $S(\theta, \phi, t)$, we assume that sunspot groups emerge in the form of regular bipolar magnetic regions (BMRs). The positive and negative regions have magnetic field distributions $B_\pm(\theta, \phi)$. The field of each new BMR is given by $S(\theta, \phi, t)=B_+-B_-$ with
\begin{equation}
B_\pm(\theta, \phi, t)=B_{\textrm{max}}\left(\frac{0.4\Delta\beta}{\delta}\right)^2\exp\left(2[1-\cos(\beta_\pm(\theta,\phi))/\delta^2]\right),
\end{equation}
where $\delta=4^\circ$. The parameter $\Delta\beta$ is determined by BMR area $A_R$ in the form of $\Delta\beta=0.45A_R^{1/2}$. $A_R$ is the sum of the sunspot group area $A_s$ and plage area $A_p$, given by the empirical formula \citep{Chapman1997}.
The parameters $\beta_\pm$ are the heliocentric angles between ($\theta, \phi$) and the central coordinates of the positive and negative polarity, ($\theta_\pm, \phi_\pm$), respectively. The coordinates of $\theta_\pm, \phi_\pm$ are determined by the colatitude $\theta$, longitude $\phi$, area $A_s$, and tilt $\alpha$ of BMRs based on the spherical geometry. The corresponding magnetic flux is determined by a single parameter, $\rm{B_{max}}$. We take $\rm{B_{max}}$ = 592 G by matching the observed total unsigned surface flux obtained from SOHO/MDI synoptic maps after rebinning them to the spatial resolution of the simulation (1$^\circ$ in both latitude and longitude) and simulations during cycle 23 \citep{Jiang2015}.  The observed inflows towards active regions \citep{Gizon2004,Gizon2008} are included in our study by multiplying tilt angles by a factor of 0.7 since past studies show that inflows decrease the amplitude of the axial dipole moment by about 30\% \citep{Martin-Belda2017}.

The dipole moment is usually used to measure the large-scale field. It is calculated by
\begin{equation}
\label{eq:dipole}
\textrm{DM}(t)=\frac{3}{2} \int_0^{180} \left\langle B\right\rangle(\theta,t)
              \cos\theta\sin\theta d\theta,
\end{equation}
where $\left\langle B\right\rangle(\theta,t)$ is the longitudinally averaged field over the solar surface. Although observations do not provide tight constraints on the values of the transport parameters, i.e., the profile of meridional flow and the turbulent diffusivity \citep{Petrovay2019b}, we do not consider the effect of varying transport parameters on the results in Section \ref{sec:results}. Figure 10 of \cite{Jiang2014a} shows that the dipole moment generated by sunspots at low latitudes and high latitudes has a weak and opposite dependence on $\upsilon(\theta)$ and $\eta$, especially within the observed range. For a whole cycle having both the high and low latitude emergence, we argue that the effects of varying transport parameters on the dipole moment are weak. This is demonstrated by \cite{Virtanen2017}, who showed that the simulated field is fairly insensitive to uncertainties in transport parameters.

Previous studies have always focused on the dipole moment at cycle minimum due to its predictive capability of the solar cycle amplitude. However, the dipole moment at solar minimum is the combined result of the cancellation of the preceding cycle's dipole moment and a generation of the new cycle's. A weaker cycle has a smaller dipole moment at the end of the preceding cycle, and vice versa \citep{Schatten1978, Cameron2015, Jiang2018}. This distracts from our understanding of the quantitative contribution of the BL mechanism to the solar cycle. To avoid this problem, we investigate the time evolution of the total dipole moment during cycle $n$, $\Delta\rm{DM}(\textit{t}) = |DM(\textit{t}) - DM_{n-1}|$, where DM$_{n-1}$ is the value at the end of cycle $n-1$. The final total dipole moment generated during cycle $n$ is $\Delta\rm{DM}_{n} = |DM_{n} - DM_{n-1}|$, which is the total dipole moment generated by all sunspot emergences during the whole cycle $n$. This definition allows us to calculate and compare the final total dipole moment $\Delta\rm{DM}_{n}$ for cycles with different populations of sunspot groups. For convenience, we set the initial magnetic field to be zero in all simulations of each individual cycle.

\begin{figure}[htbp]
\centering
\includegraphics[scale=0.9]{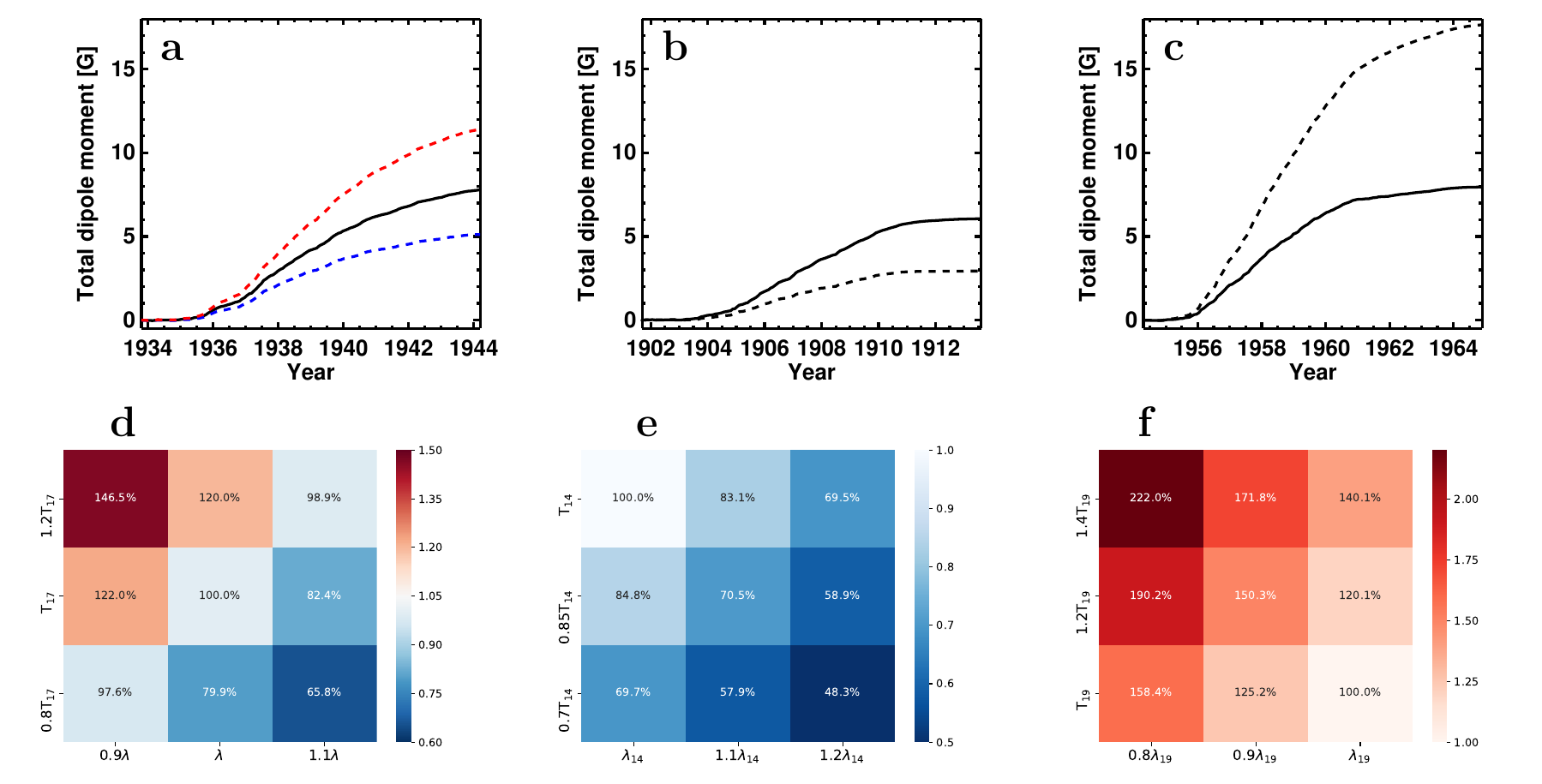}
\caption{Effects of latitudinal and tilt quenching on total dipole moment. \textbf{a,} Time evolution of the total dipole moment $\triangle\rm{DM(t)}$ with both forms of quenching for cycle 17 (solid black curve). The red (blue) dashed curve shows an extreme case scenario with a combination of a 20\% increase (decrease) of the tilt coefficient and a 10\% decrease (increase) of the observed latitudes. \textbf{b,} Time evolution of $\triangle\rm{DM(t)}$ for the weak cycle 14 with both forms of quenching in a solid black curve. The dashed curve shows the case with a combination of a 30\% decrease of the tilt coefficient and a
20\% increase of the observed latitudes. \textbf{c,} Time evolution of $\triangle\rm{DM(t)}$ for the strong cycle 19 with both forms of quenching in a solid black curve. The dashed curve shows the case with a combination of a 40\% increase of the tilt coefficient and a 20\% decrease of the observed latitudes. \textbf{d-f,} Matrix of relative total dipole moment change for different combinations of latitude $\lambda_{n}$ and tilt coefficient $T_{n}$. Values are referenced to the case with the observed latitude and tilt coefficient.}
\end{figure}

\section{Results} \label{sec:results}
\subsection{Results based on three representative cycles}

Cycles 17, 14, and 19 are the average, weakest, and strongest cycles during the RGO period. Their $S_n$ values are 197, 108, and 286, respectively. We illustrate the impact of systematic changes in the latitude and tilt of sunspot groups on $\Delta\rm{DM}(t)$ based on these three representative cycles. For the source term in the SFT simulations, the location and area of each sunspot group are from the RGO record. The tilt angle is calculated based on Eqs.(\ref{eq:alpha}) and (\ref{eq:TnSn}), without considering the tilt scatter.

For cycle 17, we use the observed $\pm10$\% variation in latitude and $\pm20$\% variation in tilt coefficient to be within the difference from other cycles over the last century. Figure 2a shows the time evolution of the simulated total dipole moment for different combinations of tilt coefficients and average latitudes. The red and blue curves are chosen to highlight the potential impact of tilt and latitudinal quenching on the dipole moment evolution. Figure 2d shows a matrix of the relative change $\Delta\rm{DM}_{n}$ for each combination of values. For cycle 14, we use the observed $20$\% increase in latitude and $30$\% decrease in tilt coefficient. For cycle 19, we use the observed $20$\% decrease in latitude and $40$\% increase in tilt coefficient. The corresponding results for the weak cycle 14 and the strong cycle 19 are presented in the 2nd and the 3rd columns of Figure 2, respectively.

The values in the matrices of the bottom panels show that latitudinal quenching has strong and similar quantitative modulations of $\Delta\rm{DM}_{n}$ as tilt quenching. Hence, latitudinal quenching plays the same role as tilt quenching in modulating cycle amplitude. The origin of the two forms of quenching has some differences. Tilt quenching is only caused by flux emergence. Latitudinal quenching is caused by both flux emergence and surface flux transport, which generates different cross-equatorial flux. \cite{Cameron2010} once included both forms of quenching in their SFT simulations. But they only addressed the role of tilt quenching in their studies.

From the top panels, we can see that although the strength of cycle 19 is about 3 times stronger than cycle 14, the difference between the final total dipolar moments they generated is small (7.73 G, 6.04 G, and 7.94 G for cycles 14, 17, and 19, respectively). This implies that strong cycles are less effective than weak cycles at generating total dipole moment. This is evidence of the role of latitudinal and tilt quenching as regulating mechanisms for the solar cycle.

Fully quantifying the effect of latitudinal and tilt quenching on total dipole moment generation by cycles with different amplitudes is not possible using observed cycles alone. This is because we have direct observations of a limited number of cycles, and each observed cycle is just one random realization of the stochastic processes determining the area, location, and tilt of sunspot groups that are part of flux emergence. The stochastic processes affect the evolution of the dipole field \citep{Jiang2014b}, which prevents us from understanding the nonlinear modulation of the solar cycle due to the systematic properties of sunspot groups. Hence, in the following subsection, we first synthesize solar cycles and then quantify the effect of the two forms of quenching.

\begin{figure}[htbp]
\centering
\includegraphics[scale=0.35]{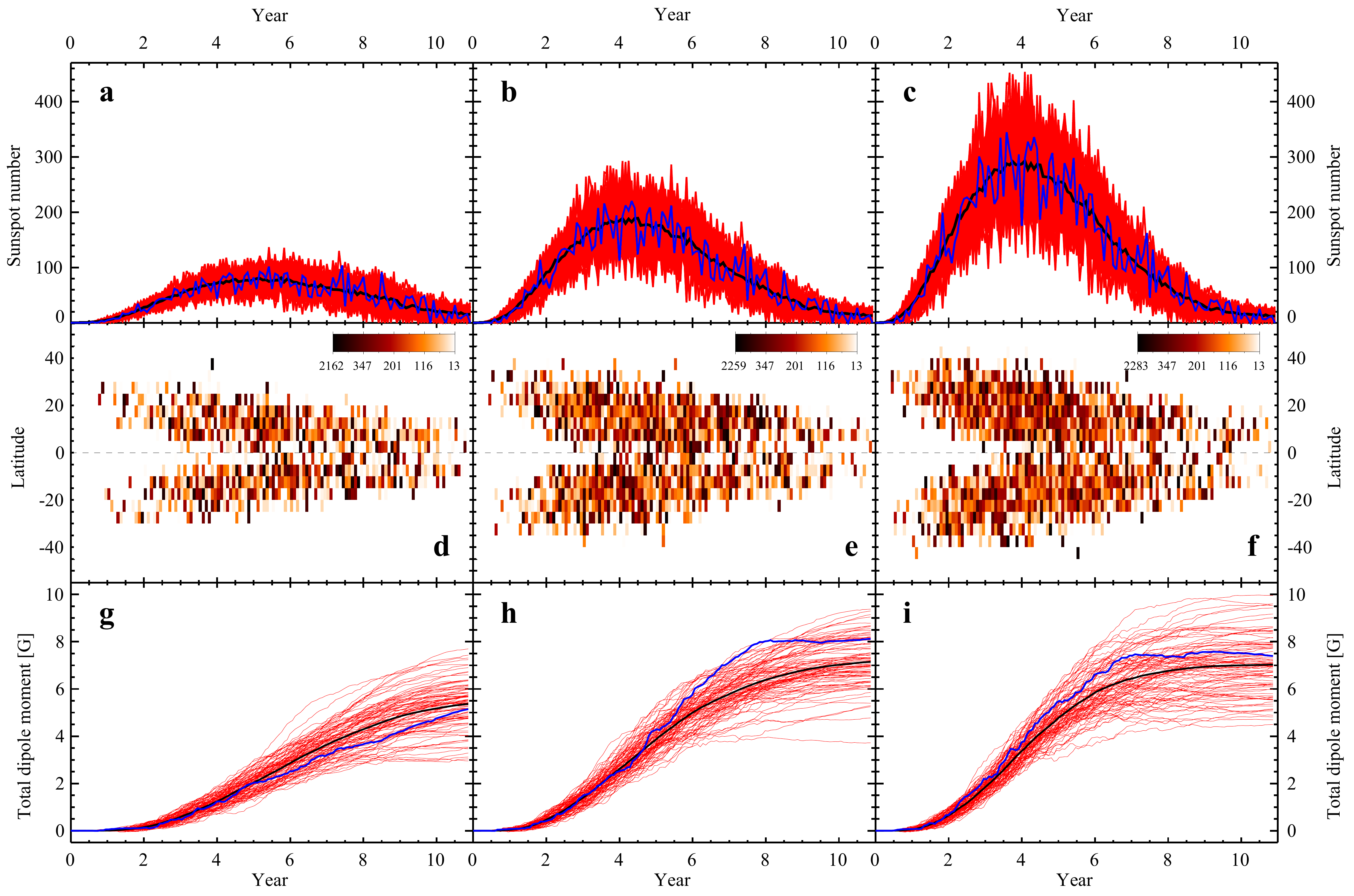}
\caption{Synthetic solar cycles of different amplitudes. Top: Time evolution of the synthetic sunspot number. Red curves show the 100 simulated random realizations. Black indicates the expected sunspot number value at each time. The blue curve marks one of the random realizations. Middle: Time evolution of the sunspot emergence latitude for the cases shown using blue curves.  The color displays the average area in units of MSH in bins of one month by 5 degrees in latitude.  Bottom: Time evolution of the total dipole moment. Each column represents results obtained using different cycle amplitudes.  From left to right $S_n$=70 (\textbf{a}, \textbf{d}, and \textbf{g}), $S_n$=180 (\textbf{b}, \textbf{e}, and \textbf{h}), and  $S_n$=280 (\textbf{c}, \textbf{f}, and \textbf{i}).}
\end{figure}

\subsection{Results based on synthetic solar cycles}

The relationships shown in Figures 1a and 1b can be connected with quantifications of the statistical distributions of sunspot group properties, as well as average solar cycle properties, to generate highly realistic synthetic solar cycles.  The properties of these cycles are determined by their amplitude. We use the method presented by \cite{Jiang2018}. The major steps to synthesize solar cycles are summarized as follows.

For the source term $S(\theta, \phi, t)$ of SFT simulations, we need the latitude, longitude, area, and tilt of emerged sunspot groups, which are approximated as BMRs, within a 1 day time interval. The first step is to get the number of daily emerged BMRs, $N_{\textrm{BMR}}(t)$, at a given time $t$. Its systematic component is determined by the cycle strength $S_n$. The random component depends on the cycle phase. The latitudinal location of each BMR also consists of the systematic and random components, both of which depend on the cycle phase and cycle amplitude. It has a symmetric latitude distribution in both the northern and southern hemispheres and a random longitudinal distribution. The synthetic areas of BMRs during a cycle obey the observed number density function of a sunspot group area. The mean areas depend on the cycle phase. The mean tilt angle is determined by Eqs.(\ref{eq:alpha}) and (\ref{eq:TnSn}). The scatter of the tilt angle depends on its area. Furthermore, we multiply a factor 0.7 to the tilt to mimic the effects of the near-surface inflows toward sunspot groups \citep{Gizon2004,Gizon2008}. For more details to synthesize solar cycles, see \cite{Jiang2011,Jiang2018}.

For each cycle amplitude, we generate 100 random realizations. The cycle amplitudes vary between 40 and 280. The roles of the non-linearities are investigated by analyzing the average behavior of the 100 random realizations. Figure 3 shows the behavior of these statistical ensembles (red), a representative example (blue), and expected values (black), for weak (left column), medium (center column), and strong cycles (right column). The top panels show the sunspot number curve, the middle panels show the distribution and area of sunspot groups in latitude and time for the representative example (butterfly diagram), and the bottom panels show the evolution of the total dipole moment. By construction, the cycle amplitude determines the number of regions, as well as their latitude and tilt. However, the expected value of the final total dipolar moment can be seen to be very weakly determined by amplitude. This has important implications for cycle variability because the expected value of the final total dipolar moment seems to be a constant property of the solar cycle.

\begin{figure}[htbp]
\centering
\includegraphics[scale=0.7]{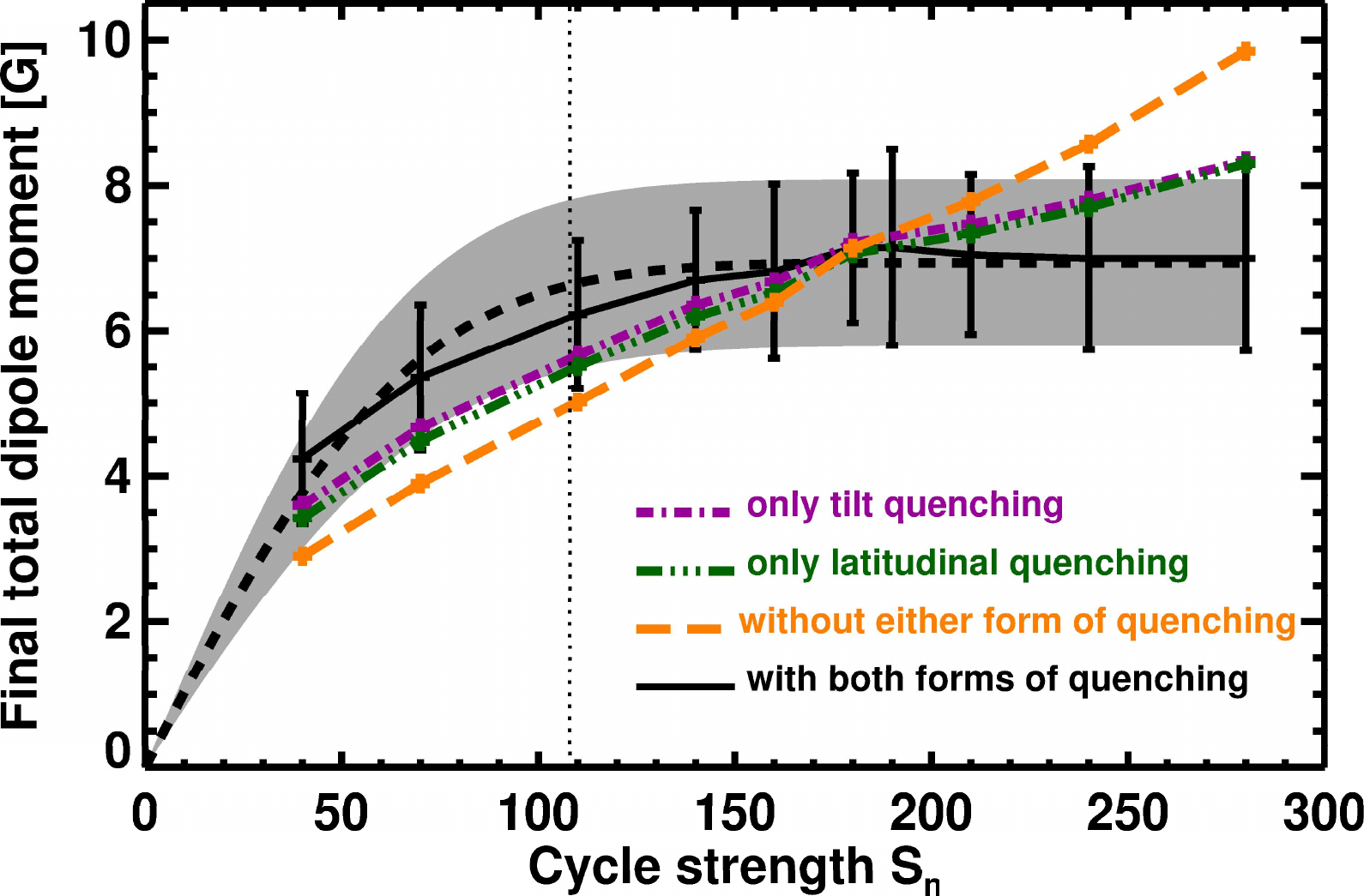}
\caption{Relation between final total dipolar moment and cycle amplitude. The black solid curve indicates the expected values from 100 SFT simulations using random sunspot group realizations including latitudinal and tilt quenching. The black dashed curve is the curve fit to the solid black curve using Eq.(\ref{eqn:SnDM}). Error bars correspond to the 1$\sigma$ standard deviation, caused by the randomness in the properties of sunspot groups. The upper and lower boundaries of the shaded region correspond to the fit to the upper and lower boundaries of the standard deviations. Green dash-triple-dot and purple dash-dot curves show the expected values for SFT simulations with only the latitudinal and tilt quenching, respectively. Orange long dashes show the expected value of SFT simulations without latitudinal nor tilt quenching. The vertical dotted line shows the minimum cycle strength to derive the empirical relations between sunspot emergence and cycle strength.}
\end{figure}

In Figure  4 we show the relationship between the final total dipolar moment $\Delta\rm{DM}_{n}$ and cycles of different amplitudes $S_n$ for simulations that: 1.\ do not have tilt or latitudinal quenching (Orange long dashes), 2.\ only have tilt quenching (Purple dash-dot), 3.\ only have latitudinal quenching (green dash-triple-dot), and 4.\ include both forms of quenching (black). We show the mean (thick lines in different colors) and standard deviation (for the case including both forms of quenching), which results from the randomness in the sunspot emergence for different simulations. Without either form of quenching the relationship between the amplitude and the expected final total dipole moment is nearly linear (i.e.,\ proportional to the strength of the simulated cycle). Including either latitudinal or tilt quenching introduces a non-linearity in which weak (strong) cycles produce a more (less) expected total dipole moment than without quenching. Including both enhances this effect into a saturation state in which normal and strong cycles tend to produce similar amounts of the total dipolar moment (regardless of their amplitudes), whereas weak cycles become even more effective in generating a total dipole moment. We expect zero $\Delta\rm{DM}_{n}$ when there is no sunspot emergence. The curve fit to the mean values of the simulations including both forms of quenching is
\begin{equation}
\Delta\textrm{DM}_{n}=6.94\mathrm{erf}\left(\frac{S_n}{75.95}\right),
\label{eqn:SnDM}
\end{equation}
which is shown in the dashed curve in Figure 4. The shaded area covers the curve fits to the upper and lower boundaries of the standard deviations.

Our results suggest that, in a normal mode of operation, the Sun has an expected, constant dipolar magnetic budget (regardless of the cycle amplitude). For strong cycles, more of this constant budget needs to be spent canceling the preceding cycle's dipole moment at cycle minimum and less on building that of the ongoing cycle; for weak cycles the converse is true. This explains the observed alternation between strong and weak cycles, i.e., the Gnevyshev-Ohl rule. It is important to note that this behavior is only expected in a statistical sense. Each cycle is still the outcome of random realizations and thus may produce statistical deviations from the Gnevyshev-Ohl rule. One such example is the weak solar cycle 24, which has been shown to deviate from an expected higher amplitude by the collective effect of sunspot groups with opposite dipole moment contribution \citep{Jiang2015}.

The non-linearities shown in Figure 4 also provide a possibility to explain the evolution of the Maunder minimum, which seems to show a sudden start and slow recovery \citep{Usoskin2017}. Sudden entry could happen when the constant dipole budget is spent almost entirely cancelling the dipole moment at the previous cycle minimum. Exit from a grand minimum state is enabled by the latitudinal and tilt quenching, which increases the relative effectiveness of weak cycles for generating the total dipole moment. \cite{Ribes1993, Munoz-Jaramillo2019, Arlt2020} showed that sunspot emergences from 1620 to 1720 occurred very close to the equator, with a trend of increasing mean latitude with the gradual recovery of the cycle amplitudes. This is consistent with the latitude's property constrained by normal cycles 12-20. This allows the Sun to gradually increase cycle strength until a normal mode of the odd-even effect kicks back in.

The bottom panels of Figure 3 present a large scatter of the final total dipole moment, which corresponds to the wide error bars in Figure 4. They result from the random components of sunspot group emergence. The impact of the randomness on the dipole moment is comparable in amplitude to that of the systematic variation. This would suggest that both the stochastic and deterministic mechanisms play roles in modulating the solar cycle. Cycles that deviate from the Gnevyshev-Ohl rules are suggested to be caused by the randomness. Except for the nonlinearities for the sudden entry into a grand minimum shown in the above paragraph, the randomness is also possible for the sudden entry \citep{Cameron2017,Nagy2017}. For extreme cycles, e.g., the strongest cycle 19, the nonlinear mechanisms could play the dominant role. For moderate cycles, e.g, cycle 17, randomness could statistically play the dominant role. Since we aim to propose the observable nonlinearities of the solar cycle and their effects in this paper, we do not evaluate the importance of the nonlinear mechanisms compared with the randomness in modulating the solar cycle here.

\section{Conclusions}
\label{sec:summary}
This work demonstrates that modulation of sunspot group latitude and tilt by  cycle amplitude plays a vital role in regulating solar variability. In particular, that systematic changes in latitudes of emerging bipolar active regions have a similar nonlinear effect on the solar dynamo as tilt quenching, something that has long been overlooked in studies of the solar cycle. The combined effect of latitude and tilt quenching makes the BL mechanism a self-regulating process. This prevents the unbounded growth of the cycle amplitude and favors the recovery from states of very low activity. These feedback mechanisms are backed by observational evidence. The nonlinearities we suggest show that the BL mechanism is a nonlinear process in essence. The nonlinearities act as the amplitude-limiting mechanisms for BL-type dynamos. Dynamical nonlinearities, i.e., the backreaction of Lorentz forces on the flows in the form of the Malkus-Proctor effect \citep{Malkus1975, Rempel2006}, $\Lambda$-quenching \citep{Kichatinov1993, Kichatinov1994} and so on, have been suggested as other possible means of regulating the solar cycle. The nonlinearities inherent in the BL mechanism are distinct from these dynamical nonlinearities, which are not directly observable. Different types of nonlinearities could work together to modulate the solar cycle. Their relative importance remains an open question.

The latitudinal dependence of sunspot emergence on the solar cycle results from the destabilization and rise of flux tubes, which form from the toroidal field. These correspond to Process I of the BL mechanism given in the Introduction. The whole process leading to the observed properties of sunspot emergence, i.e., Process I, is still an open question. But we can observe its final output, i.e., the sunspot emergence, which provides the source of surface magnetic flux. The surface flux transport processes, i.e., Process II of the BL mechanism, due to meridional flow and turbulent diffusion introduce the nonlinear modulation of the final total dipole moment. The final total dipole moment is determined by the cross-equatorial flux, which depends on the latitudinal location of sunspot emergence and is well-described by a Gaussian profile \citep{Jiang2014b,Petrovay2020}. A lower latitude emergence contributes a much more final dipole moment because much more leading polarity flux can be diffused across the equator. The property of the flux transport combined with the property of sunspot emergence, i.e., stronger cycles having higher latitude emergence, generates the latitudinal quenching.

The anti-correlation between the cycle strength and the tilt coefficient given by Figure 1b has not been widely accepted by the community because the anti-correlation is weak and the cycle 19 point contributes most to the weak anti-correlation. Concerning this issue, we argue as follows. First, numerical simulations of rising flux tubes through the convection zone support the anti-correlation since stronger flux tubes rise more rapidly and suffer less distortion by the Coriolis force during their rise \citep{DSilva1993,Caligari1995,Weber2013}. Except for this support, we here suggest another explanation for the anti-correlation. \cite{Martin-Belda2017} showed that inflows towards active regions are a potential non-linear mechanism capable of modulating the solar cycle. Inflows reduce the axial dipole moment at the end of the cycle by ~30\% with respect to the case without inflows in cycles of moderate activity. This ratio varies by ~9\% from very weak cycles to very strong cycles. We suggest that inflows might be responsible for the anti-correlation between the tilt coefficient and the cycle strength. The ~9\% variation of the ratio due to inflows might correspond to our tilt quenching. Secondly, the tilt angle is noisy in essence because of buffeting by convective turbulence during the rise of the flux tube \citep{Weber2013}. So a strong anti-correlation is not expected. Thirdly, tilt angle of cycles 21-24 from a new dataset supports the anti-correlation (Jiao et al., in prep.). The two extreme cycles, i.e., the strongest cycle 19 and weakest cycles 24, have the smallest and largest tilt coefficients, respectively. Even if the anti-correlation does not exist, and hence the tilt quenching does not work on the Sun, the latitudinal quenching could still be a nonlinear mechanism to modulate the solar cycle.

Now we come back to suggest both forms of quenching operate on the Sun. Whether the solar dynamo is quasi-periodic or chaotic is a fundamental question to understand solar cycle variability and predictability. The answer depends on the strength of the non-linear mechanisms. \cite{Charbonneau2005} suggested that for different ranges of the dynamo number, the non-linearity inherent in BL-type dynamo models can lead to a stable cycle, cycle doubling, a chaotic solution, and other different dynamical behaviours. We can give the expression of the non-linearity we present here and further the 1D iterative map based on this study. An evaluation of the non-linearity and a comparison with the properties of the long-term variability of solar activity will be investigated in a subsequent paper. The other ingredient that is essential to understanding solar cycle variability and predictability is the randomness in the sunspot emergence. \cite{Cameron2017, Lemerle2017} showed that it is the stochasticity that dominates the observed variability of solar activity on a wide range of timescales. The relative effects of the systematic and random properties on modulating the solar cycle remain to be evaluated.

Although the proposed non-linear mechanisms of the solar cycle are based on surface magnetic flux evolution, and thus not a dynamo model in itself, we would argue that this method is not inferior to any attempts based on current BL dynamo models. We still do not have a proper understanding of the subsurface physics, e.g., the distribution and flux emergence process of the toroidal field \citep{Spruit2011, Karak2014}. By our method we can bypass the unknown subsurface physics and put an emphasis on the final result caused by the subsurface physics. The final result, i.e., the systematic dependence of sunspot emergence on the cycle strength, can be directly measured. Furthermore, the correlation between the polar field at cycle minimum and the subsequent cycle strength \citep{Munoz-Jaramillo2012, Jiang2018} supports the linear process from the poloidal field to the toroidal field. This helps to justify our method in identifying the nonlinear mechanism of the solar cycle based on surface magnetic flux evolution.

\acknowledgments
I thank Andr\'{e}s Mu\~{n}oz-Jaramillo, Robert Cameron, Manfred Sch\"{u}ssler, and the anonymous referee for insightful comments that help to improve the work. I also thank Andr\'{e}s Mu\~{n}oz-Jaramillo for language corrections on the original version of the article. The sunspot records are courtesy of WDC-SILSO, Royal Observatory of Belgium, Brussels. This research was supported by the National Natural Science Foundation of China through grant Nos. 11873023 and 11522325, the Fundamental
Research Funds for the Central Universities of China, Key Research Program of Frontier Sciences of CAS through grant No. ZDBS-LY-SLH013, the B-type Strategic Priority Program of CAS through grant No. XDB41000000, and the International Space Science Institute Teams 474 and 475.

\vspace{5mm}

%


\bibliography{Nonlinerity}{}
\bibliographystyle{aasjournal}



\end{document}